\documentclass[fleqn,usenatbib]{mnras}

\usepackage{newtxtext,newtxmath}

\usepackage[T1]{fontenc}
\usepackage{ae,aecompl}
\usepackage{graphicx}
\usepackage{booktabs}
\usepackage{multirow}
\usepackage{xspace}
\usepackage{comment}
\usepackage{siunitx}
\usepackage{overpic}
\DeclareSIUnit{\eV}{eV}
\DeclareSIUnit{\keV}{keV}
\DeclareSIUnit{\parsec}{pc}
\DeclareSIUnit{\erg}{erg}
\DeclareSIUnit{\gauss}{G}
\DeclareSIUnit{\year}{yr}
\usepackage{hyperref}
\def\msun{{\rm M}_{\odot}}

\def\ibh {OGLE-2011-BLG-0462}
\def\bh {BLG-0462}
\def \degmark{^\circ}
\def \mdot {\dot M}

\newcommand{\cha}{\textit{Chandra}\xspace}

\usepackage{natbib} 

\title[blz]{
Deep Chandra X-ray observation of the isolated black hole \ibh\ }

\author[ ]{Sandro Mereghetti$^{1}$,  Lara Sidoli$^{1}$, Gabriele Ponti$^{2,3,4,5}$, Aldo Treves$^{4}$
\\
$^{1}$ INAF, Istituto di Astrofisica Spaziale e Fisica Cosmica Milano, via A.\ Corti 12, I-20133 Milano, Italy\\
$^{2}$     INAF, Osservatorio Astronomico di Brera, Via E. Bianchi 46, Merate (LC), I-23807, Italy\\
$^{3}$     Max-Planck-Institut für extraterrestrische Physik, Giessenbachstrasse, 85748, Garching, Germany\\
$^{4}$       Universit\`a dell'Insubria, Dipartimento di Scienza e Alta Tecnologia, Via Valleggio 11, I-22100 Como, Italy\\
$^{5}$    Como Lake Center for Astrophysics (CLAP)  \\
}

\date{Accepted 21 July 2025. Received  11 July 2025 ; in original form  4 June 2025}

\pubyear{2025}

\begin{document}
\label{firstpage}
\pagerange{\pageref{firstpage}--\pageref{lastpage}}
\maketitle

%
%
 
  \begin{abstract}
\ibh\ is an isolated black hole of $\sim7\msun$ at a distance of 1.5 kpc identified thanks to the astrometric microlensing technique. It is the first specimen discovered of the large population of $\sim10^8$ stellar-mass black holes that are believd to wander in the Galaxy. Electromagnetic radiation  powered by accretion from the interstellar medium is expected  from \ibh, but has not been detected at any wavelength. We present the results of a deep pointed observation   with the \cha\ satellite that provides an upper limit of 3$\times10^{29}$ erg s$^{-1}$ on the luminosity of \ibh\ in the 0.5-7 keV energy range. This is about one order of magnitude below the previous limit obtained from shallower observations that serendipitously covered the sky position of this black hole. Our results are briefly compared with models of the source and with the X-ray upper limits for candidate isolated black holes and  black holes in wide binary systems.
\end{abstract}
   \begin{keywords}
Gravitational lensing: micro -- stars: black holes, individual: OGLE-2011-BLG-0462 -- accretion  
   \end{keywords}

   \maketitle
%

\section{Introduction} 

A large number (10$^7$-10$^9$) of isolated black holes (IBHs) are expected be present  in the Milky Way. Unlike black holes in interacting binary systems, that can be revealed through their accretion-powered X-ray emission (e.g. \citealt{2006ARA&A..44...49R}), IBHs are difficult to detect.   
Although several authors considered the possibility of revealing isolated neutron stars and IBHs by detecting their electromagnetic emission powered by accretion from the interstellar medium (e.g. \citealt{1991A&A...241..107T,1993A&A...277..477C,1998ApJ...495L..85F,2018MNRAS.477..791T,2021ApJ...922L..15K,2025arXiv250620711M}) none has been identified up to now. 
An alternative way to find these  very elusive ``dark'' objects is through their gravitational lensing effect on the light of background stars \citep{1986ApJ...304....1P,1992ApJ...392..442G}.

In the last few years,  encouraging results in the search for galactic IBHs have been obtained thanks to the addition of milliarcsecond precision astrometry to gravitational microlensing observations (e.g. \citealt{2024A&A...692A..28K}).  When the deviations in the apparent position of the gravitationally magnified star are measured, it is possible to reduce the degeneracy between the relative velocities and distances of the star and gravitational lens, and thus to better constrain the mass of the latter. 
Several dark lenses with masses consistent with that of a neutron star or a black hole have been found with photometric microlensing \citep{2002ApJ...579..639B,2002MNRAS.329..349M}, and for a few candidates it was also possible to obtain accurate astrometry with Gaia and the Hubble Space Telescope.
Among them,  \ibh\ is the most convincing case of being the first-discovered stellar-mass IBH in the Milky Way. 

\ibh\ (also known as MOA-2011-BLG-191, \bh\ in the following) was independently discovered  by two groups searching for microlensing events in the  direction of the Galactic bulge.
A lens mass of   $M_L=7.1\pm1.3$ $\msun$ was derived by   \citet{2022ApJ...933...83S}, while \citet{2022ApJ...933L..23L}, with  a  different analysis of nearly the same data set, found two possible solutions: one implying a black hole with lower mass ($M_L=3.8 \pm 0.6$ $\msun$)  and the other one consistent also with a neutron star ($M_L=2.15^{+0.67}_{-0.54}$  $\msun$).  
The reasons for this discrepancy were thoroughly investigated by \citet{2022ApJ...937L..24M} and \citet{2023ApJ...955..116L}. The lens masses found in these reanalysis   ($M_L=7.88\pm0.82$ $\msun$  and $M_L=6.0^{+1.2}_{-1.0}$  $\msun$, respectively) are in much better agreement and clearly rule out a neutron star nature for the gravitaional lens.  In the following we will adopt the most recent parameters of \bh\  derived by \citet{2025ApJ...983..104S}, i.e. a black hole mass of $M_L=7.15\pm0.83$ $\msun$, a transverse velocity of $51.1\pm7.5$ km s$^{-1}$, and a distance of $1.52\pm0.15$ kpc.

A first search for X-ray emission from \bh\ based on short serendipitous observations found in public data archives was presented in \citet{2022ApJ...934...62M}, where we also reported upper limits on the X-ray flux for other isolated black holes and neutron star candidates found with microlensing. Here we present the results of a deep dedicated observation of \bh\ obtained with the \cha\ observatory.

\section{Observations and Data Analysis}
\label{ODA}

\begin{figure*}
\centering
\includegraphics[width=13cm]{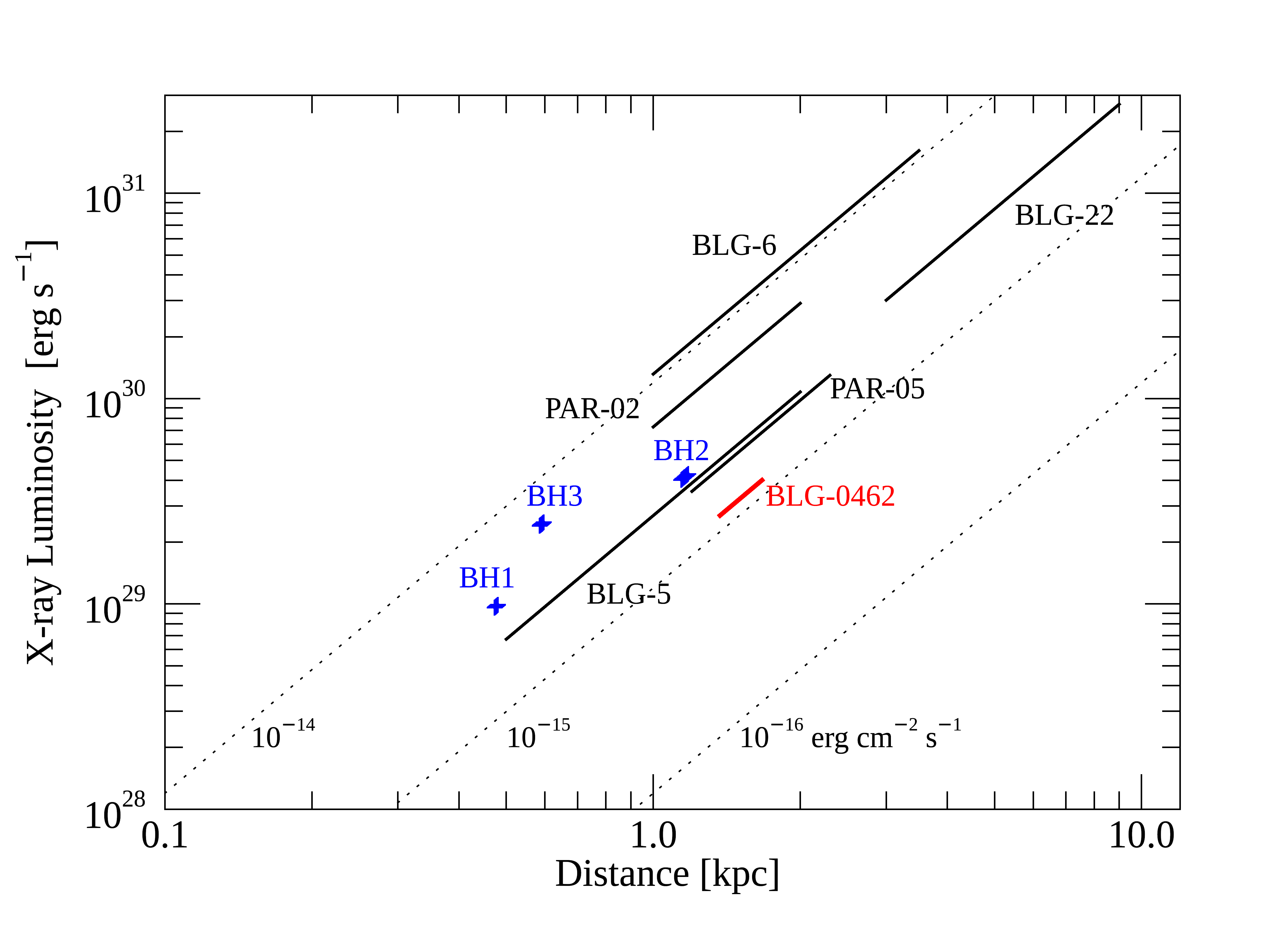}
\caption{Upper limits on X-ray luminosity for BLG-0462 (red) and other candidate IBHs found with microlensing (black). The length of the lines reflects the uncertainty on the source distances. The upper limits for three BHs in non-interacting binaries are indicated in blue. Dashed lines indicate the flux as a function of distance.}
\label{Fig:1}
\end{figure*}
 
\begin{table*}
\caption{Dormant (candidate) black holes in the Galaxy}
\begin{center}
 \hspace{-30mm}
\begin{tabular}{lcclccl}
\hline
Name & Mass & Distance    & Flux$^a$     &   Companion & Orbital period & References$^b$ \\
          & [$\msun$]  &  [kpc] &  [erg cm$^{-2}$ s$^{-1}$]     &  star & [yr] &  \\
\hline
 \ibh\     &   7.15$\pm$0.83 & 1.52$\pm$0.15 &  $<1.2\times10^{-15}$  & no & --& [1]\\
MACHO-96-BLG-5   &  6$^{+10}_{-3}$   &  0.5$-$2  &  $<2.2\times10^{-15}$ & no &--&  [2,3,4]\\  
MACHO-98-BLG-6   &  6$^{+7}_{-3}$    &  1$-$3.5   &  $<1.1\times10^{-14}$   & no & --& [2,4]\\ 
MACHO-99-BLG-22   & 7.5$\pm$3   &  3$-$9 &  $<2.8\times10^{-15}$  & no &--& [5,4] \\  
OGLE3-ULENS-PAR-02 &  11.9$^{+4.9}_{-5.2}$ &  1.3$^{+0.7}_{-0.3}$   &  $<6.08\times10^{-15}$   & no & -- & [6,4] \\
OGLE3-ULENS-PAR-05 &   6.7$^{+3.2}_{-2.7}$  &  1.6$^{+0.7}_{-0.4}$  &  $<2.11\times10^{-15}$ & no & -- & [6,4] \\  
Gaia BH1  &   9.62$\pm$0.18 & 0.477$\pm$0.004 &  $<3.58\times10^{-15}$  & G dwarf & 0.51 & [7,8] \\ 
Gaia BH2  &  8.9$\pm$0.3 & 1.16$\pm$0.02 &  $<2.58\times10^{-15}$ & red giant & 3.5& [7,9]  \\ 
Gaia BH3  &  32.70$\pm$0.82  & 0.591$\pm$0.006  &  $<5.86\times10^{-15 ~c}$ & G giant 
& 11.6  & [10,11] \\
\hline
\end{tabular}
\label{table:BH}
\end{center}
\raggedright
$^a$ In the 0.5-7 keV range, corrected for the absorption.

$^b$ References: 
[1]  \citet{2025ApJ...983..104S},
[2] \citet{2002ApJ...579..639B},
[3]  \citet{2006ApJ...651.1092N},
[4]  \citet{2022ApJ...934...62M},
[5] \citet{2005ApJ...633..914P},
[6] \citet{2020A&A...636A..20W},
[7]  \citet{2024PASP..136b4203R},
[8] \citet{2023MNRAS.518.1057E},
[9] \citet{2023MNRAS.521.4323E},
[10]  \citet{2024ApJ...973...75C},
[11] \citet{2024AA...686L...2G}.

$^c$ Upper limit at apastron.
\end{table*}

We observed \ibh\ with   \cha\   from  2024 September 7 at 14:49 UT  to September 8 at 06:30 UT.  The observation  was carried out with the 
 ACIS-I instrument in VFAINT mode  and  provided a livetime exposure of  53.5 ks.  
 Contrary to the   ACIS-I images analysed in \citet{2022ApJ...934...62M},  in which \bh\ was serendipitously observed  at large off-axis angles (from 5 to $\sim$9 arcmin), the new observation was pointed at the coordinates of the target:  
 R.A. = 17$^{h}$51$^{m}$40.2082$^{s}$,   
 Dec. = $-29\degmark$53$'$26.502$''$ (J2000).  
 
 No photons at the position of \ibh\ were detected in this deep observation.  Following the same procedure used in \citet{2022ApJ...934...62M}, we derived an upper limit of  6.5$\times$10$^{-5}$ counts s$^{-1}$ on the source count rate in the 0.5--7 keV energy range (95\% confidence level,  based on \citealt{1991ApJ...374..344K}). This is a factor 8 lower than the previous limit.
 Assuming a power law spectrum with photon index $\alpha$=2 and interstellar absorption $N_H=10^{21}$ cm$^{-2}$, this count rate corresponds to a limit on the observed (unabsorbed) 0.5-7 keV flux of 1.0$\times$10$^{-15}$ erg cm$^{-2}$ s$^{-1}$  
 (1.2$\times$10$^{-15}$ erg cm$^{-2}$ s$^{-1}$). For a   distance of  1.52 kpc 
 the  upper limit on the X-ray luminosity in the same energy range is 3.3$\times$10$^{29}$ erg  s$^{-1}$, which corresponds to a fraction $3\times10^{-10}$ of the Eddington luminosity, $L_{Edd}$, of a black hole of 7.15 $\msun$.

\section{Discussion and conclusions}
\label{sec:conc}

An isolated black hole wandering in the interstellar medium is expected to gravitationally capture some matter and emit electromagnetic radiation from the resulting accretion flow.
The simplest estimate of the accretion rate onto a black hole of mass $M$ moving with velocity $V$ in a medium of density $n$ is provided by the Bondi-Hoyle-Lyttleton formalism

\begin{equation}
\mdot = 4\pi n m_p    \frac{(GM)^2}{(V^2 + c_s^2)^{3/2}}  \lambda   ~~{\rm  g~s^{-1}},
\label{eq-mdot}
\end{equation}
  
\noindent
where $c_s$ is the sound speed, G is the gravitational constant, and $m_p$ is the proton mass. The parameter $\lambda \lesssim1$ accounts for the fact that not all the matter within the Bondi-Hoyle radius is accreted onto the black hole. 
Recent magnetohydrodynamical simulations indicate typical values of $\lambda \sim0.01-0.1$ \citep{2022A&A...660A...5B,2023ApJ...950...31K,2025ApJ...978..148G,2025PhRvD.111h3025K}. In the following we normalise the equations to $\lambda_{0.1}=\lambda/0.1$.

 Adopting the values for the mass and velocity of  \bh\ given above, and neglecting $c_s$, which is much smaller than $V$, equation \ref{eq-mdot} gives an accretion rate of

\begin{equation}
\mdot =   1.42\times10^{10} ~ n \lambda_{0.1}  ~~~~~{\rm  g~s^{-1}}.
\label{eq-mdot2}
\end{equation}

\noindent
The corresponding luminosity is

\begin{equation}
L_x = \eta \mdot c^2 =  1.27\times10^{31} ~  n \lambda_{0.1} ~\eta   ~~~~~{\rm  erg~s^{-1}},
\label{eq-mdot2}
\end{equation}

\noindent
where $\eta$ is the efficiency of conversion of gravitational energy into electromagnetic radiation in the observed X-ray band. 

The upper limit on the  luminosity obtained with \cha\ implies an X-ray  efficiency 

\begin{equation}
\eta    < \frac{2.6\times10^{-2}}{n \lambda_{0.1}}.
\end{equation}

\noindent
This limit is consistent with the low X-ray radiative efficiency expected in theoretical models for black holes accreting at greatly sub-Eddington rate (see, e.g., \citealt{2014ARA&A..52..529Y}, and references therein).

\citet{2025arXiv250301172K} showed that the accretion flow onto \bh\ probably proceeds through a magnetically arrested disc (MAD). 
In the MAD accretion regime the magnetic field is sufficiently strong and ordered to affect the dynamics of the accretion flow and the main energy dissipation is magnetic reconnection rather than turbulence. 
Under the MAD scenario, 
these authors predict for \bh\  an X-ray luminosity of 
 
\begin{equation}
L_x \sim 10^{28} ~ \bigg( \frac{n \lambda}{0.1~{\rm  cm^{-3}}}\bigg) \bigg(\frac{\eta f_e \epsilon_{NT} \epsilon_{dis}}{0.1~0.3~0.05}\bigg)  ~~{\rm  erg~s^{-1}},
\label{eq-mdot2}
\end{equation}

\noindent 
where $f_e$ is the electron heating fraction, $\epsilon_{\rm NT}$ is the non-thermal particle production efficiency and $\epsilon_{\rm dis}$ is the dissipation energy fraction. 
For typical values of these parameters, eq.~\ref{eq-mdot2}  predicts an X-ray flux of about $4\times10^{-17}$ erg cm$^{-2}$ s$^{-1}$, well below the sensitivity of our \cha\ observation. 
A similar prediction, based on the accretion model of  \citet{2005A&A...440..223B}, was reported by \citet{2022AstBu..77..223C}.

It is  interesting to compare the X-ray luminosity limits of \bh\ with those obtained for a few other (candidate) black holes of stellar mass, which are likely in regimes of low accretion rate (see Table~\ref{table:BH}). Five of the objects listed in the table, besides \bh , are candidate IBHs found with gravitational microlensing. Although they can be considered as promising candidates for being IBHs, this has not been confirmed with certainty, and their mass and distances are much less constrained than in the case of \bh . We list in Table~\ref{table:BH} also  three black holes found in  binary  systems. Their masses have been precisely measured through spectroscopic observations of their companion stars and also their distances are well constrained. These black holes are in wide orbits with long periods and might, in principle, accrete from the stellar winds of their companions. 
For example, assuming Bondi-Hoyle-Lyttleton accretion and based on the estimated (although quite uncertain) properties of its companion's wind,  BH3 at apastron\footnote{The orbital eccentricity of BH3  is e=0.73} is expected to accrete at a rate 
%
%
$4.4\times10^{11} < \dot{M} < 4.4\times10^{13}$  g s$^{-1}$ 
\citep{2024ApJ...973...75C}.

The constraints on the X-ray emission for the (candidate) black holes of Table~\ref{table:BH} are plotted in Fig.~\ref{Fig:1}, where it can be seen that the  flux limit for \bh\ is the lowest one.   However, a more relevant quantity to consider is the X-ray luminosity which depends on the distance.  In this respect,   BLG-5  could provide more constraining limits on the accretion efficiency, but only if its BH nature is confirmed and its distance found to be of only a few hundred parsecs. The other candidate IBHs have worse flux limits than \bh\ and they are at larger, and more unconstrained  distances.    In terms of luminosity, the confirmed black holes BH1 and BH3 found in Gaia binaries have slightly smaller limits \citep{2024PASP..136b4203R,2024ApJ...973...75C}. As discussed by these authors, they provide strong constraints on the accretion efficiency because it is possible to estimate the properties of the stellar winds from which they accrete and, in addition, they have precise distance measurements.
Observations with future X-ray satellites such as $NewAthena$ \citep{2025NatAs...9...36C} can provide a sensitivity down to $\sim10^{-17}$ erg cm$^{-2}$ s$^{-1}$, two orders of magnitude below the current best limits, thus measuring, or strongly constraining, the X-ray luminosity of these black holes. However, at such low luminosity levels, the likely presence of X-ray emission from the companion stars might complicate the interpretation of the results for the black holes in binary systems. 
The IBHs found through gravitational microlensing will not be affected by this problem, due to the much weaker X-ray contamination expected from the lensed stars. For example, the background star lensed by \bh\ is a late-type subgiant at a distance of 7.1 kpc \citep{2025ApJ...983..104S}.

Up to now we considered   accretion  as the only form of energy production  in \bh .  However, if the black hole were rotating, especially if the specific angular momentum  were close to 1, it could act as a unipolar generator (e.g., \citealt{2015PASJ...67...89O,2024Ap.....67..420B}). At present, the issue of how the related luminosity should compare with that from the accretion of the interstellar medium is essentially unexplored. In any case, our upper limit on the X-ray luminosity could become a significant constraint of this kind of modelling.
It is obvious that a direct detection of \bh\  in any spectral band would be extremely valuable. 
Prospects for detection of IBHs in the radio band have been discussed by \citet{2005MNRAS.360L..30M,2013MNRAS.430.1538F}  in the context of accretion powered models  and based on the empirical correlation between   radio and X-ray luminosities observed in X-ray binaries. More recently, \citet{2019MNRAS.488.2099T} estimated the radio luminosity of IBH considering the radiatively inefficient flows expected at low accretion rates and predicted that several tens will be detected with the upcoming Square Kilometer Array.
The presence of relativistic outflows makes   IBH also potentially interesting sources of non-thermal high-energy emission (e.g. \citealt{2012MNRAS.427..589B,2025ApJ...985..251K}), but it should be noted that the relatively poor angular resolution achievable at gamma-ray energies hampers a firm identification of such objects (which might already be present among unidentified gamma-ray sources). 

Both the accretion and unipolar scenarios consider the possibility of electromagnetic showers of very low energy content. However, because of the dimensions of the emitting region and of the expected values of the magnetic field, it is possible that, as in the case of pulsars, the radio emission exhibit some kind of coherence (see \citet{2018A&A...613A..61G} for a recent discussion on this topic). Contrary to the case of pulsars, the phenomenon could appear sporadically.  This would suggest long systematic observations  in the low frequency radio band. An advantage for the identification of the radio counterpart of \bh\  would be that its position is very well known.


\section*{Acknowledgements}
We thank the anonymous referee for her/his useful comments.
We acknowledge financial support from  INAF through the 2022 grant for fundamental research. GP acknowledges support from the European Research Council (ERC) under the European Union’s Horizon 2020 research and innovation program HotMilk (grant agreement No. 865637), support from Bando per il Finanziamento della Ricerca Fondamentale 2022 dell’Istituto Nazionale di Astrofisica (INAF): GO Large program and from the Framework per l’Attrazione e il Rafforzamento delle Eccellenze (FARE) per la ricerca in Italia (R20L5S39T9).
This research has made use of data obtained from the \cha\ Data Archive and software provided by the \cha\ X-ray Center (CXC) in the application package CIAO.
 
 \section*{Data availability}
All the data used in this article are available in public archives.\\



\bsp	
\label{lastpage}
\end{document}